\documentclass[twocolumn]{aastex631}
\usepackage{amsmath}
\usepackage{bm}
\usepackage{subfigure}
\usepackage{multirow}
\usepackage{makecell}

\definecolor{db}{HTML}{191586}
\definecolor{vi}{HTML}{5E1111}

\newcommand{\Ek}{\mathrm{E}}
\newcommand{\ct}{C_\mathrm{t}}

\newcommand{\bn}{{\boldsymbol\nabla}}
\newcommand{\uw}{\bm u }
\newcommand{\ft}{\bm f_\mathrm{t}}
\newcommand{\edr}{E_\mathrm{dr}}
\newcommand{\ikd}{\mathrm{Im}\left[k_2^2(\omega)\right]}
\newcommand{\qpo}{Q_\omega'}
\newcommand{\qp}{Q'}

\newcommand{\ep}{\epsilon_\mathrm{t}}
\newcommand{\epOm}{\epsilon_\Omega}

\begin{document}

\title{Tidally-excited inertial waves in stars and planets: \\exploring the frequency-dependent and averaged dissipation with nonlinear simulations}

\author[0000-0002-8561-0769]{Aurélie Astoul}
\affiliation{School of Mathematics, University of Leeds,\\ Leeds LS2 9JT, UK
}

\author[0000-0003-4397-7332]{Adrian J. Barker}
\affiliation{School of Mathematics, University of Leeds,\\ Leeds LS2 9JT, UK
}

%\collaboration{20}{(AAS Journals Data Editors)}

%\author{F.X Timmes}
%\affiliation{Arizona State University}
%\affiliation{AAS Journals Associate Editor-in-Chief}

%\altaffiliation{AASTeX v6+ programmer}

%\affiliation{AAS Director of Publishing}
%\affiliation{American Astronomical Society \\
%1667 K Street NW, Suite 800 \\



\begin{abstract}
We simulate the nonlinear hydrodynamical evolution of tidally-excited inertial waves in convective envelopes of rotating stars and giant planets modelled as spherical shells containing incompressible, viscous and adiabatically-stratified fluid. This model is relevant for studying tidal interactions between close-in planets and their stars, as well as close low-mass star binaries. We explore in detail the frequency-dependent tidal dissipation rates obtained from an extensive suite of numerical simulations, which we compare with linear theory, including with the widely-employed frequency-averaged formalism to represent inertial wave dissipation. We demonstrate that the frequency-averaged predictions appear to be quite robust and is approximately reproduced in our nonlinear simulations spanning the frequency range of inertial waves as we vary the convective envelope thickness, tidal amplitude, and Ekman number. Yet, we find nonlinear simulations can produce significant differences with linear theory for a given tidal frequency (potentially by orders of magnitude), largely due to tidal generation of differential rotation and its effects on the waves. Since the dissipation in a given system can be very different both in linear and nonlinear simulations, the frequency-averaged formalism should be used with caution. Despite its robustness, it is also unclear how accurately it represents tidal evolution in real (frequency-dependent) systems.
\end{abstract}


\section{Introduction} \label{sec:intro}
Tidal interactions play an important role in driving spin-orbit evolution in planetary and close stellar binary systems \citep[e.g.][]{Z2013,O2014}. A key mechanism in low-mass stars and giant planets with convective envelopes is tidal excitation of inertial waves (hereafter IWs) if the perturbed body rotates sufficiently rapidly. IWs are essentially incompressible disturbances in rotating fluids restored by Coriolis forces, which propagate in approximately neutrally-stratified convection zones. These waves are excited by tidal forcing if the tidal frequency $\omega$ satisfies $|\omega|\leq 2\Omega$, where $\Omega$ is the rotational angular velocity. The properties of these waves, and their contribution to tidal dissipation (hence to tidal torques and spin-orbit evolution), depend strongly on the internal structure of the body \citep[e.g.][varying with stellar mass, age, rotation and metallicity]{M2015,GB2017,BG2017,B2020} and physical mechanisms at play \citep[like differential rotation or magnetism,][]{BR2013,GB2016,W2018,LO2018,AM2019}. 

Most prior work studying tidal IWs has employed linear theory of a non-magnetised viscous fluid in a uniformly rotating spherical shell \citep[e.g.][]{OL2004,OL2007,O2009,GL2009,RV2010}. Linear theory is valid if tidal amplitudes are sufficiently small for nonlinearities to be unimportant. However, some close-in planets (such as Hot Jupiters) and stars in close binary systems may have sufficiently large tidal amplitudes for important nonlinear effects that could considerably alter tidal dissipation rates. We have therefore started to explore in detail the nonlinear evolution of tidally-excited IWs in convective envelopes of stars and giant planets in \citet[][hereafter AB22]{AB2022}, building upon \cite{FB2014} and \cite{Barker2016}.

For astrophysical modelling of tidal spin-orbit evolution in planetary and stellar systems, many authors have employed the linear frequency-averaged IW dissipation following \cite{O2013}. This is obtained by applying an impulsive tidal forcing to obtain an ordinary differential equation describing the wave-like response of the fluid, which can be straightforwardly solved to determine the frequency-averaged IW dissipation. For a piece-wise homogeneous stellar model, this provides simple analytical expressions for tidal dissipation rates and quality factors due to IWs, as employed in many prior studies \citep[e.g.][and many others]{M2015,BM2016,G2017,B2020}. More recently, this has been computed using realistic stellar models \cite[][not piece-wise homogeneous]{B2020}, applied to model planetary orbital migration \citep{L2021}, and to explain circularisation of solar-type stellar binaries \citep{B2022}. Despite its simplicity and wide usage to model stellar and planetary populations, the robustness of frequency-averaged dissipation predictions from linear theory have yet to be verified in nonlinear simulations, or those incorporating turbulent convection, differential rotation or magnetic fields \citep[though a magnetised homogeneous shell in linear theory has been studied by][]{LO2018,W2018}.

We build upon AB22 and model nonlinear tidally-forced IWs in neutrally-stratified spherical shells in three dimensions, representing convective envelopes of low-mass stars (from M to F spectral types, with masses ranging from $0.4$ to $1.4M_\odot$) or giant gaseous planets. We explore the parameter space covering the entire range of wave frequencies for IWs (between $-2\Omega$ and $+2\Omega$ in the fluid frame), radial aspect ratios $\alpha$ (ratio of inner to outer radii i.e.~shell thickness, to model various stars and planets), and Ekman numbers $\Ek$ (ratio of viscous to Coriolis forces), so as to determine the validity of linear theoretical predictions, including the frequency-averaged dissipation measure commonly applied to model astrophysical systems. We introduce our model in \S~\ref{model}, discuss our results in \S~\ref{scan}, present applications to observed astrophysical systems in \S~\ref{astro}, and finally we conclude in \S~\ref{conc}.
%%%%%%%%%%%%%%%%%%%%%%%%%%%%%%%%%%%%%%%%%%%%%%%
 \section{Modelling nonlinear tidal inertial waves}\label{model}
We adopt the model of AB22 to study tidally-forced IWs with frequency $\omega$, in an adiabatically-stratified, incompressible, viscous fluid, in a spherical shell with constant density $\rho$ and initial uniform rotation $\Omega$.
In the frame rotating at the rate $\boldsymbol{\Omega}=\Omega\boldsymbol{e}_z$ of the tidally-perturbed body (with spherical coordinates $(r,\theta,\varphi)$), the wavelike response, with velocity field $\uw$ and pressure $p$, satisfies the momentum and continuity equations (AB22):
\begin{equation}
\begin{aligned}
  \partial_t\bm u+(\uw\cdot\bn)\uw+2\bm\Omega\wedge\bm u
   &=-\frac{\bn p}{\rho}+\ft+\nu\Delta \bm u, \\
   \bn\cdot\bm u& =0.
\end{aligned}
\label{setw}
\end{equation}
Here $\bm f_\mathrm{t}=-2\bm\Omega\wedge\bm u_\mathrm{e}$ is the Coriolis acceleration on the non-wavelike tidal flow $\bm u_\mathrm{e}$ (equivalent to the conventional equilibrium tide here), which acts as an effective body force to excite (inertial) wavelike tides \citep[e.g.][]{O2013}. The volume-integrated tidal dissipation rate is $D_\nu=-\left<\rho\nu\bm u \cdot\Delta \bm u\right>$ (where $\langle \cdot\rangle$ denotes a volume integral), with $\nu$ assumed to be a constant effective viscosity modelling the action of convection on wavelike tides.
This assumption is motivated in particular by \citet[][and references therein]{DB2020a} and \citet{VB2020a,VB2020b} albeit for equilibrium tides, though the treatment of convective effects in this way is still debated \citep[][]{T2021,BA2021}. The time-averaged $D_\nu$ balances the corresponding tidal power $P_\mathrm{t}=\langle\bm u \cdot \bm f_\mathrm{t}\rangle$ in a steady state.

We adopt the planetary/stellar radius $R$, and $\Omega^{-1}$, as our units of length and time, respectively. The non-wavelike tidal flow $\bm u_\mathrm{e}$ is described by a set of time-dependent equations in linear theory\footnote{But it is perfectly maintained on the timescale of our simulations, meant to be short compared with tidal evolution timescales (see also Appendix \ref{time}).}: the momentum, continuity and Poisson equations, and leading-order quasi-hydrostatic equilibrium \citep[plus boundary conditions of a tidally-perturbed free surface at $r=R$ and impenetrability on $r=\alpha R$,][]{O2013}. We consider only the tidal component with harmonic degree and order $l=m=2$ (using spherical harmonics $Y_l^m$), which is (one of) the dominant component(s) for asynchronous (eccentricity) tides in a coplanar two-body system, so
\begin{equation}
    \bm u_\mathrm{e}= \mathrm{Re}\left[
    \mathrm{i}\omega\bm\nabla X(r,\theta,\varphi)\,\mathrm{e}^{-\mathrm{i}\omega t}\right],
\end{equation}
with
\begin{equation}
    X(r,\theta,\varphi) =\frac{C_{\rm t}}{2(1-\alpha^5)} \left[ r^{2} + \frac{2}{3}\alpha^5 r^{-3} \right]Y^2_2(\theta,\varphi),
\end{equation}
where $\ct=\left(1+\mathrm{Re}\left[k^2_2\right]\right)\epsilon$ is the tidal forcing amplitude, related to the real part of the quadrupolar Love number $\mathrm{Re}\left[k^2_2\right]$ (subsequently abbreviated to $k_2$) and the tidal amplitude parameter $\epsilon=(M_2/M_1)(R/a)^3$ ($M_2$ and $M_1$ are masses of the perturber and perturbed body, respectively, the latter has radius $R$ and orbital semi-major axis $a$). 

We introduce several quantities to analyse our simulations. The energy in the differential rotation $\edr$ triggered by nonlinear IW interactions, integrated over the volume $V$ of the shell, is defined as \cite[see also:][]{T2007,FB2014,AB2022}:
\begin{equation}
    \edr=\frac{\rho}{2}\langle \left[\langle u_\varphi\rangle_\varphi-\delta\Omega~ r\sin\theta\right]^2\,\rangle,
\end{equation}
where $\langle\cdot\rangle_\varphi$ denotes a $\varphi$-average, and  
\begin{equation}
    \delta\Omega=\frac{1}{V}\left\langle\frac{u_\varphi}{r\sin\theta}\right\rangle,
\end{equation}
is the mean rotation rate of the fluid in the $\Omega$-frame, where  $\Omega^*=\Omega+\delta\Omega$ is the modified rotation rate in the inertial frame\footnote{$\delta \Omega$ is nonzero not from tidal synchronisation -- we do not observe this gradual process directly since we study short snapshots in the evolution of the system -- but because of the induced differential rotation.}. The zonal flow strength depends on the tidal forcing amplitude $C_\mathrm{t}$, and different scaling laws can be derived for low and high $C_\mathrm{t}$  (as done in Appendix \ref{scal}).

For modelling tidal evolution we wish to compute the dissipation rate (or tidal power). A useful quantity is the modified tidal quality factor $Q'$, which is proportional to the ratio between the maximum stored tidal energy and the time-averaged dissipation rate \citep{G1963}. To obtain this from simulations we compute the dimensionless tidal dissipation rate $D_\nu$, related to the imaginary part of the Love number $\mathrm{Im}\left[k^2_2\right]$ according to \citep{O2013,LO2018}:
\begin{equation}
    \hat{D}_\nu(\omega)=\frac{5R\Omega}{8\pi G}|A|^2\omega\,\ikd,
    \label{eq:dimdiss}
\end{equation}
where $\hat{D}_\nu(\omega)=\rho\Omega^3R^5 D_\nu(\omega)$ is the dimensional tidal dissipation rate, $A$ is the tidal potential amplitude proportional to $GM_2R^2/a^3=\epsilon \omega_d^2 R^2$,
and $\omega_\mathrm{d}=\sqrt{GM_1/R^3}$ is the characteristic dynamical frequency. This leads to 
\begin{equation}
\label{eq:im2}
    \left|\ikd\right|=\frac{6}{5}\frac{\ep^2}{\ct^2}
    \frac{D_\nu (\omega)}{|\omega|},
\end{equation}
and hence the frequency-dependent modified tidal quality factor (for $l=m=2$)
\begin{equation}
\label{qfactor}
    \qpo\equiv\frac{3}{2|\mathrm{Im}\left[k^2_2(\omega)\right]|}=\frac{5}{4}\frac{\ct^2}{\ep^2}
    \frac{|\omega|}{D_\nu (\omega)},
\end{equation}
where $\ep=\epOm(1+k_2)$, and $\epOm=\Omega/\omega_\mathrm{d}$ is small when the body is slowly rotating (which justifies neglecting centrifugal forces). The frequency-averaged $\mathrm{Im}\left[k^2_2\right]$ is defined as 
\begin{equation}
\Lambda\equiv\int_{-\infty}^{+\infty}\ikd\,\frac{\mathrm{d}\omega}{\omega}, \label{eq:Lamb}
\end{equation}
with corresponding tidal quality factor $Q'=3/(2\Lambda)$.

% Since $k_2$ is often unknown because it depends on the internal structure of the body and its change due to tidal stress, the modified tidal quality factor is often preferred $Q'=3Q/(2k_2)=3/\left(2\mathrm{Im}\left[k^2_2\right]\right)$, noting that the hydrostatic value for an homogeneous fluid body is $k_2=3/2$ \citep[][]{L1911,O2014}.

\begin{figure*}
    \centering
    \subfigure[$\alpha=0.3, \ct=5\cdot 10^{-2}, \mathrm{E}=10^{-5}$]{\includegraphics[width=\textwidth]{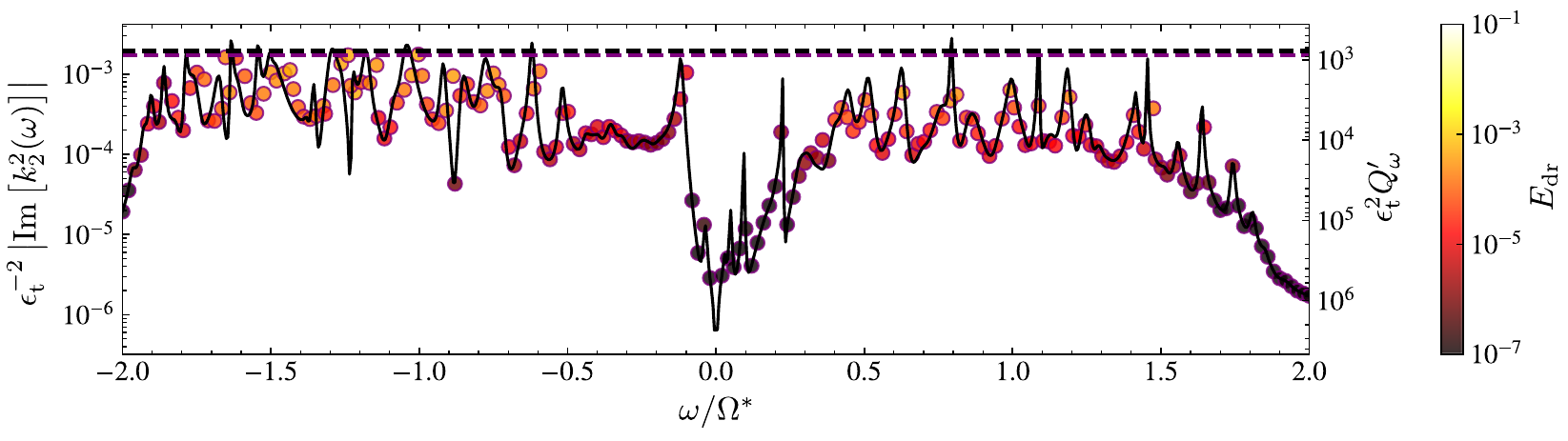}\label{fig:diss0p3}}
    \subfigure[$\alpha=0.5, \ct=5\cdot 10^{-2}, \mathrm{E}=10^{-5}$]{\includegraphics[width=\textwidth]{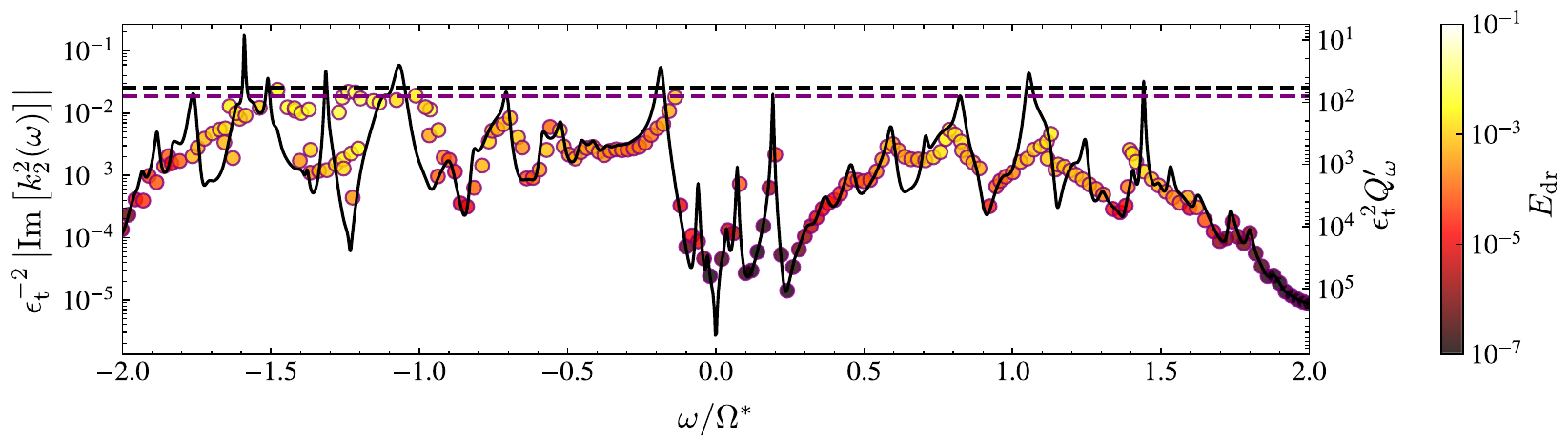}\label{fig:diss0p5}}
    \caption{Tidal dissipation from simulations as a function of frequency in the fluid frame $\omega/\Omega^*$. We show the rescaled imaginary part of the Love number $\ep^{-2}|\ikd|$ (left $y$-axis) and the tidal quality factor $\ep^2\qpo$ (right $y$-axis). Colours represent energy in the differential rotation $\edr$. Black solid and dashed lines show the frequency-dependent and frequency-averaged linear predictions given by Eqs.~(\ref{eq:im2}) and (\ref{eq:Lambda1}), respectively, while the purple dashed line is the frequency-averaged nonlinear tidal dissipation from Eq.~(\ref{eq:Lamb}). Top: radial aspect ratio $\alpha=0.3$ and tidal forcing amplitude $\ct=5\cdot10^{-2}$. $\edr$ is maximal when $\omega=-1.24$ with $\edr\approx6.2\cdot10^{-4}$, and minimal when $\omega=0.12$ with $\edr\approx4.6\cdot10^{-10}$. Bottom: Same for a thicker envelope with $\alpha=0.5$. $\edr$ is maximal when $\omega=-1.16$ with $\edr=9.5\cdot10^{-3}$, and minimal when $\omega=2.0$ with $\edr=1.7\cdot10^{-8}$.}
    \label{Fig1}
\end{figure*}

\begin{figure*}
    \centering
    \subfigure[$\alpha=0.7, \ct=10^{-2}, \mathrm{E}=10^{-5}$]{\includegraphics[width=0.9\textwidth]{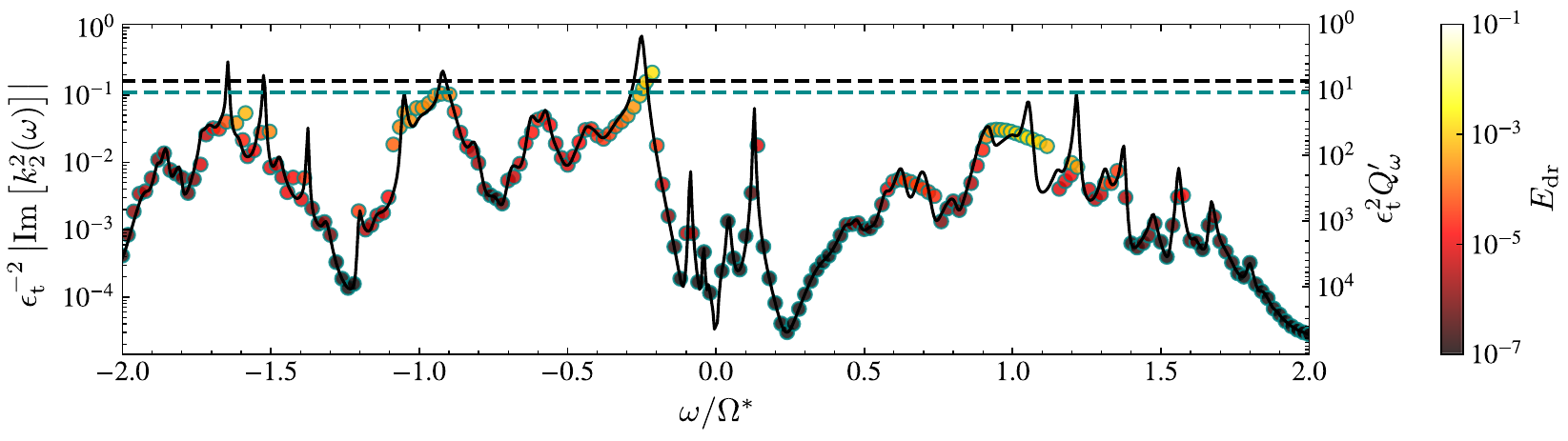}\label{0p7lowCt}}
    \subfigure[$\alpha=0.7, \ct=5\cdot10^{-2}, \mathrm{E}=5\cdot10^{-5}$]{\includegraphics[width=0.9\textwidth]{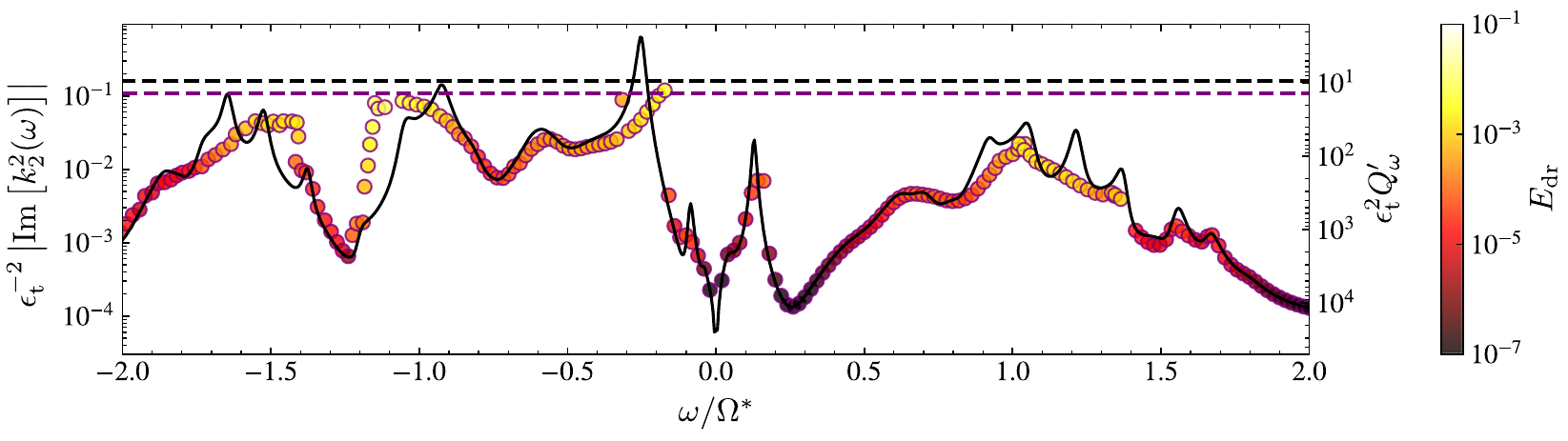} \label{0p7highE}}
    \subfigure[$\alpha=0.7, \ct=5\cdot10^{-2}, \mathrm{E}=10^{-5}$]{\includegraphics[width=0.9\textwidth]{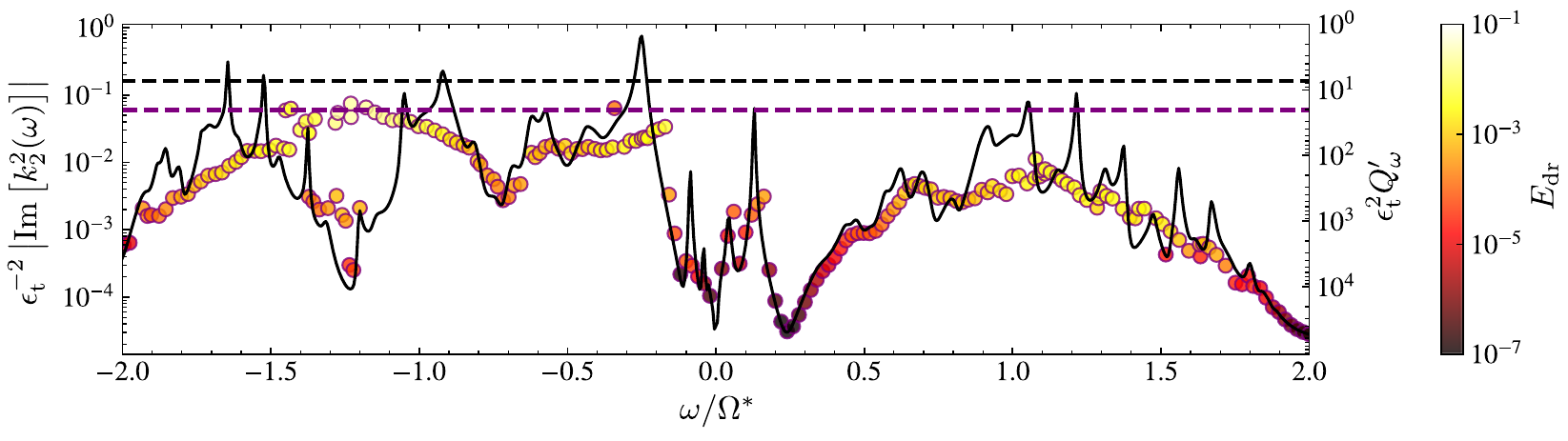} \label{0p7highCt}}
   \subfigure[$\alpha=0.7, \ct=5\cdot10^{-2}, \mathrm{E}=10^{-5}$]{\includegraphics[width=0.89\textwidth]{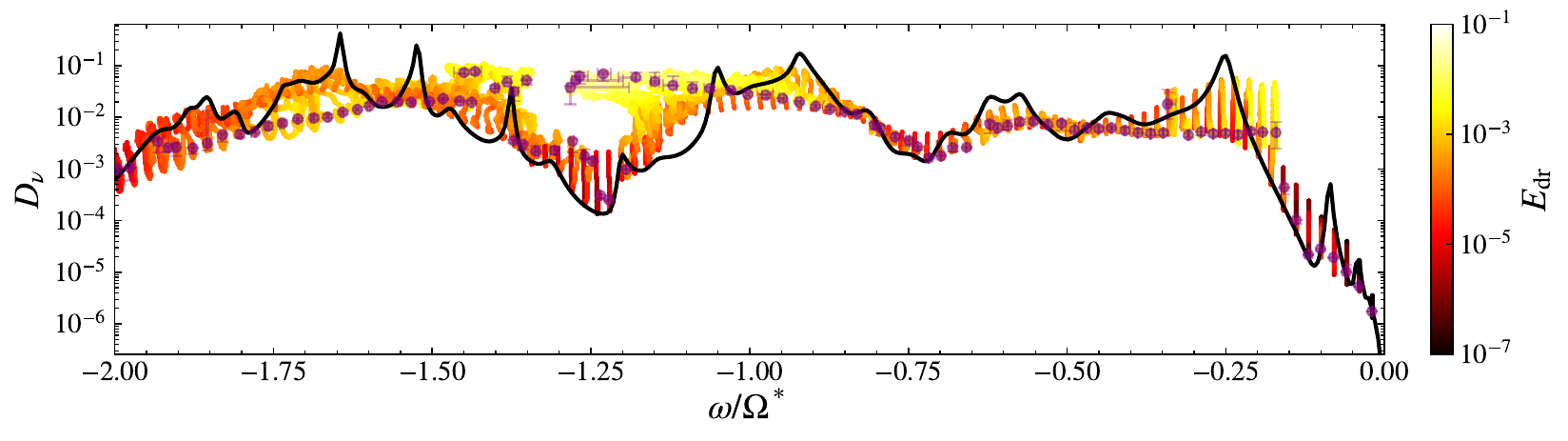}\label{fig:dnualpa0p7_neg}}
    \caption{Same as Fig.~\ref{Fig1} but for a solar-like convective envelope with $\alpha=0.7$. \textit{(a):} $\ct=10^{-2}$. Differential rotation is maximal when $\omega=-0.22$ with $\edr=1.6\cdot10^{-3}$, and minimal when $\omega=0.24$ with $\edr=7.0\cdot10^{-11}$. \textit{(b):} $\ct=5\cdot10^{-2}$ and $\Ek=5\cdot10^{-5}$. Differential rotation is maximal when $\omega=-1.06$ with $\edr=1.2\cdot10^{-2}$, and minimal when $\omega=0.26$ with $\edr=2.8\cdot10^{-8}$. 
    \textit{(c):} $\ct=5\cdot10^{-2}$ (and $\Ek=10^{-5}$ again). $\edr$ is maximal when $\omega=-1.14$ with $\edr=3.1\cdot10^{-2}$, and minimal when $\omega=0.24$ with $\edr=4.1\cdot10^{-8}$. %32% 
    $D_\nu$ can differ by up to three orders of magnitude at a given frequency depending on $\edr$. \textit{(d):} Same as (c) but coloured curves instead show evolution of $D_\nu$ and $\omega/\Omega^*$ during each simulation until (in most cases) an averaged steady state is reached. Errorbars indicate variations in the last $2000\Omega^{-1}$ compared to the final value (purple dots).
    }
    \label{fig:diss0p7}
\end{figure*}

In an incompressible fluid body containing a rigid core, \cite{O2013} derived a simple expression for the frequency-averaged tidal dissipation\footnote{Eq.~(\ref{eq:Lambda1}) is equivalent to Eq.~(113) in \cite{O2013} without assuming the same core and envelope density.}:
\begin{equation}
    \Lambda=\frac{16\pi}{63}\frac{\alpha^5}{1-\alpha^5}\ep^2.
    \label{eq:Lambda1}
\end{equation}
Since the value of $\ep$ is specific to a given stellar/planetary system and is usually not well-known, we rescale Eqs.~(\ref{eq:im2})--(\ref{eq:Lambda1}) to remove it in Figs.~\ref{Fig1}, \ref{fig:diss0p7}, and \ref{fig:diss0p9}.
%%%%%%%%%%%%%%%%%%%%%%%%%%%%%%%%%%%%%%%%%%%%%%%
%%%%%%%%%%%%%%%%%%%%%%%%%%%%%%%%%%%%%%%%%%%%%%%
%%%%%%%%%%%%%%%%%%%%%%%%%%%%%%%%%%%%%%%%%%%%%%
\section{Nonlinear tidal dissipation: results} \label{scan}
We solve Eqs.~(\ref{setw}) with the 3D pseudo-spectral code MagIC\footnote{https://magic-sph.github.io/} (version 5.10), adopting stress-free impenetrable conditions on spherical boundaries at $r=\alpha R$ and $r=R$. We mostly fix the Ekman number $\mathrm{E}=\nu/(\Omega R^2)$
to $10^{-5}$ unless otherwise stated, and we vary $\omega\in [-2\Omega,2\Omega]$ for different $\ct$ and radial aspect ratios $\alpha\in\{0.3,0.5,0.7,0.9\}$. The frequency range is scanned using $200$ equally-spaced values for each $\alpha$ and $\ct$ (and $\mathrm{E}$), which is a fair compromise between having a sufficiently good coverage of the frequency spectrum to obtain a robust frequency-averaged dissipation\footnote{With this frequency spacing, the error upon integrating to obtain the frequency-averaged value is a few percent in linear calculations. The discrepancy is harder to quantify in nonlinear simulations since we do not know the true value a priori, and exploring this is a primary aim.}, but with a reasonable total computational cost. In most simulations, we set the maximum spherical harmonic degree to $l_\mathrm{max}=85$ ($256$ longitudinal and $128$ latitudinal grid points, respectively) and use $n_r=97$ radial (Chebyshev) grid points, though higher radial and horizontal resolutions are used when necessary to ensure convergence. 
In some simulations we used $n_\varphi=512$ ($l_\mathrm{max}=170$) to guarantee adequate horizontal resolution (at least 3 orders of magnitude difference in the energy spectrum between the peak and the highest resolvable wavenumbers). Simulations are usually run for times $t\gtrsim 5000\,\Omega^{-1}$, which is usually sufficient to reach an averaged steady state. We use a CNAB2 scheme with an adaptive timestep satisfying a CFL condition no larger than $\mathrm{dt}=10^{-2}\,\Omega^{-1}$ to guarantee adequate time resolution. 

We start our investigation with a thick convective envelope ($\alpha=0.3$) relevant to model main sequence M-type stars with masses $\sim0.35M_\odot$ \citep[e.g.][]{AP2016,GB2017}, and giant planets possessing solid or stably stratified fluid cores. This also models young low-mass stars with $M\gtrsim0.4M_\odot$ during the pre-main sequence, probably the dominant phase for IW tidal dissipation \citep[e.g.][]{B2022}. For forcing amplitude $\ct=10^{-2}$, we find tidal dissipation rates in nonlinear simulations to exhibit very small departures from linear predictions when scanning the full frequency range (not shown; probably due to weak differential rotations with $E_\mathrm{dr}\lesssim10^{-6}$). We display in Fig.~\ref{fig:diss0p3} the rescaled dissipative quantities $\ep^{-2}\ikd$ and $\ep^2\qpo$ and their associated frequency-averaged values for a higher tidal amplitude $\ct=5\cdot10^{-2}$, spanning the full range of frequencies for IWs. The frequency-dependent (reddish bullets indicate final steady state values) and averaged (purple dotted line) nonlinear tidal dissipation rates do not depart significantly from linear predictions in black (solid and dashed lines, respectively). This is likely to result from the moderate zonal flow strengths triggered, as quantified by $E_\mathrm{dr}\lesssim 6.2\cdot 10^{-4}$. The differences are strongest close to ``resonant peaks" where dissipation $D_\nu$ is maximised (note that division by $\omega$ in $\Lambda$ enhances contributions from small $\omega$), where zonal flows are strongest (see also AB22). The discrepancy between linear and nonlinear frequency-averaged values 
is approximately ten percent here. 
 
Now turning to thicker envelopes with $\alpha=0.5$, relevant for M stars with masses $\sim 0.4\,M_\odot$, 
or for giant planets with extended dilute strongly stably stratified fluid cores \citep[$\alpha\sim0.6$ is inferred for Saturn,][]{MF2021} -- we find a similar order of magnitude for the discrepancy between linear predictions and nonlinear simulations with $\ct=10^{-2}$ to those with $\alpha=0.3$ and $\ct=5\cdot10^{-2}$. 
For strong tidal forcing with $\ct=5\cdot10^{-2}$, zonal flows are stronger as we can see from the larger values of $\edr$ in Fig.~\ref{fig:diss0p5}. The frequency-dependent dissipation now departs strongly from the linear prediction by several orders of magnitude for certain frequencies, especially between $-1.8$ and $-0.9$, and between $0.7$ and $1.5$, near to resonant peaks in $D_\nu$. However, the resulting frequency-averaged value deviates by less than $30\%$ from the linear prediction.

In Fig.~\ref{fig:diss0p7}, we show results for a solar-like envelope ($\alpha=0.7$), for both weak $\ct=10^{-2}$ (panel \ref{0p7lowCt}) and strong $\ct=5\cdot10^{-2}$ (panels \ref{0p7highCt} and \ref{fig:dnualpa0p7_neg}) forcing, and also for a larger viscosity $\Ek=5\cdot10^{-5}$ (\ref{0p7highE}). We now observe a larger tidal amplitude ($\ct=5\cdot10^{-2}$) or lower viscosity ($\Ek=10^{-5}$) leads to stronger departures from linear predictions, by up to three orders of magnitude at certain frequencies. This is again especially apparent in mid-range negative and positive frequencies (and around $\omega=-0.2$, probably associated with excitation of a global Rossby mode, seen more clearly for $\alpha=0.9$). We also observe strong nonlinear effects have flattened the ``peaky spectrum" predicted by linear theory by smoothing resonant peaks of enhanced dissipation. This is evidenced more clearly in Fig.~\ref{fig:dnualpa0p7_neg}, showing the evolution of $D_\nu$ versus Doppler-shifted frequency $\omega/\Omega^*$ (indicating $\edr$ by the colours), following each simulation from its static initial state towards an averaged steady state (corresponding to  Fig.~\ref{0p7highCt}). This shows a strong correlation between $\edr$, the evolution of $D_\nu$ (that departs further from linear predictions when zonal flows are strong) and $\omega/\Omega^*$. $\edr$ can be as large as $0.1$, indicating very strong tidally-driven differential rotation that may be comparable with or even larger than convectively-driven differential rotation for large $\ct$.

\begin{figure*}
    \centering
    \includegraphics[trim=0cm 0.7cm 0.5cm 0cm, clip,width=0.32\textwidth]{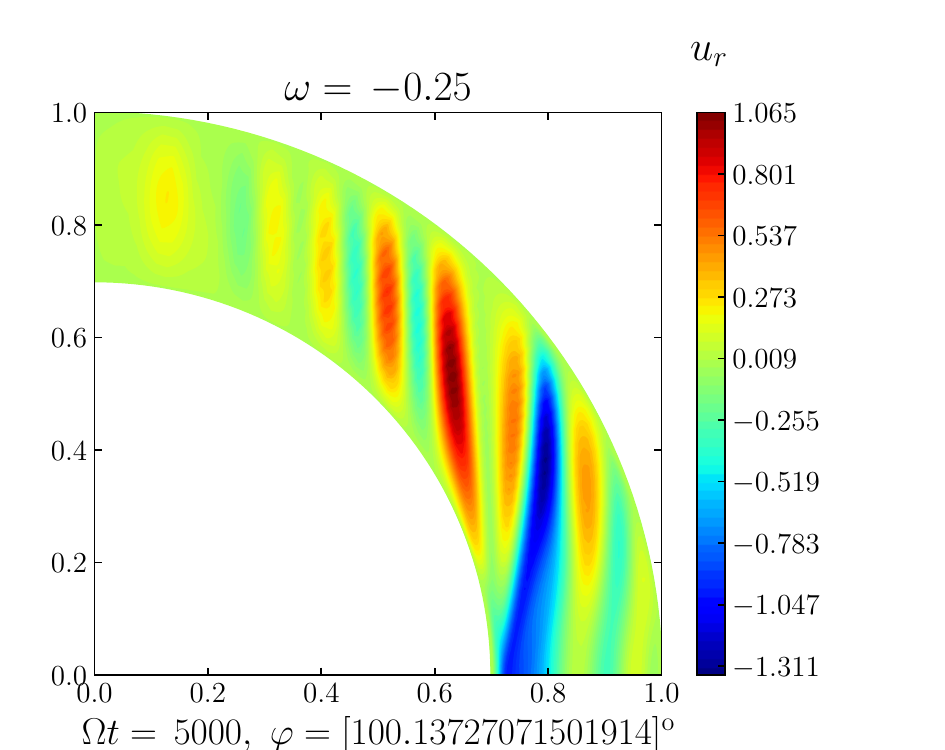}
    \includegraphics[trim=0cm 0.7cm 0.5cm 0cm, clip,width=0.32\textwidth]{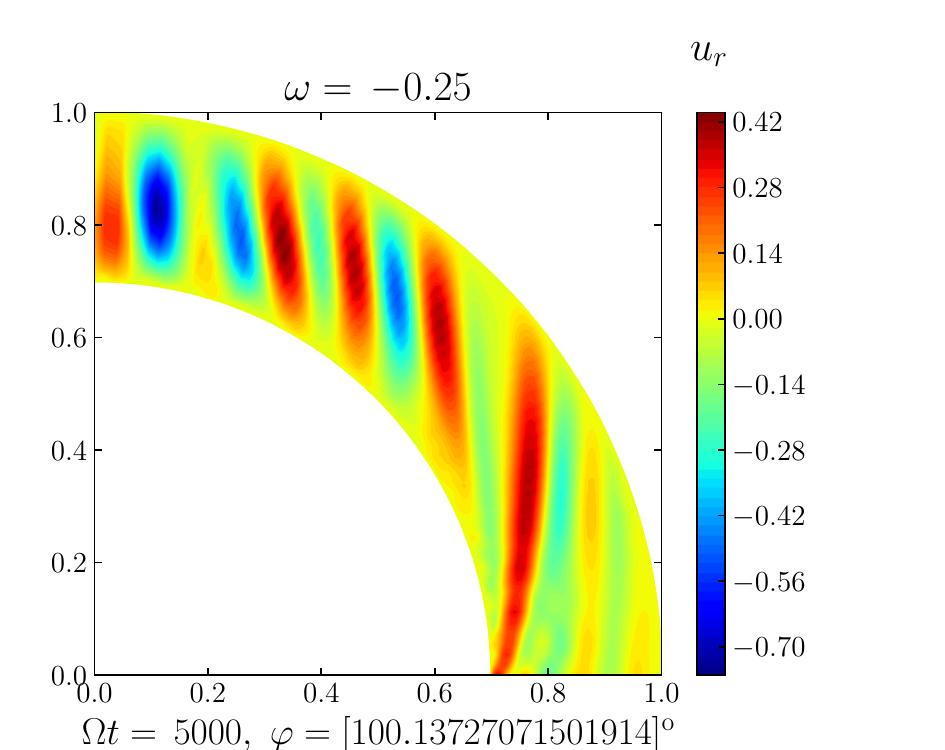}
    \includegraphics[trim=0cm 0.7cm 0.5cm 0cm, clip,width=0.32\textwidth]{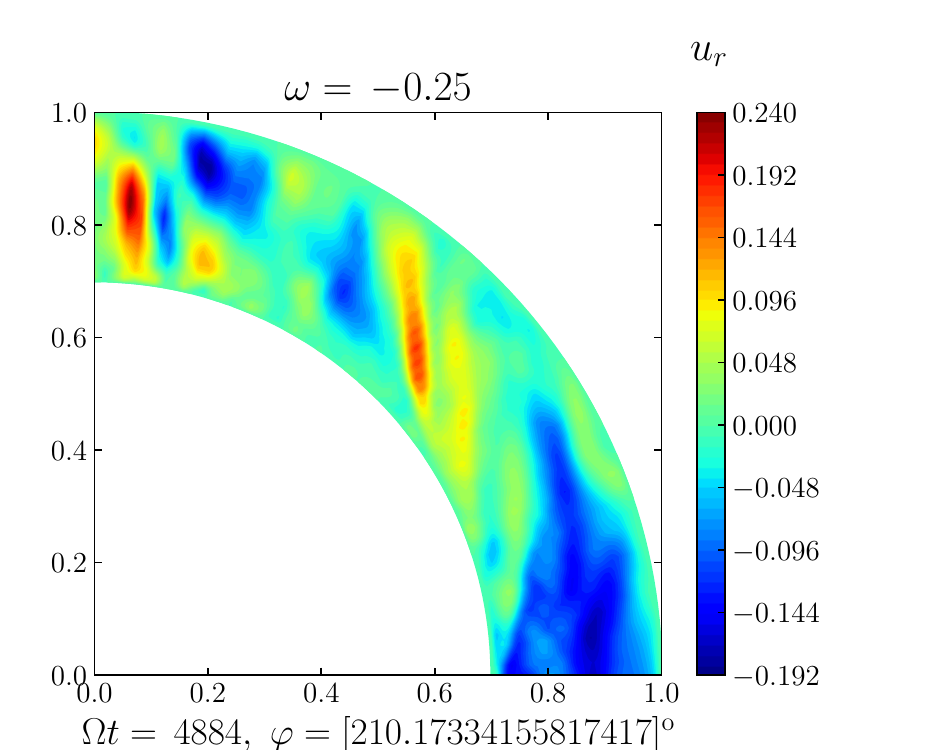}
    \caption{Meridional slice of radial velocity $u_r$ in one quadrant  with $\omega=-0.25$ and $\alpha=0.7$ (and $\mathrm{E}=10^{-5}$). \textit{Left:} $\ct=10^{-3}$ (approximately linear). Linear calculations with the spectral code LSB \citep{VR2007} are indistinguishable since nonlinearities are negligible in this case. \textit{Middle:} $\ct=10^{-2}$. \textit{Right:} $\ct=5\cdot10^{-2}$. Note the paradoxical decrease in $u_r$ when increasing $\ct$.}
    \label{fig:ur_omm0p25}
\end{figure*}
%%%%%%%%%%%%%%%%%%%%%%%%%%%%%%%
\begin{figure*}
    \centering\includegraphics[width=\textwidth]{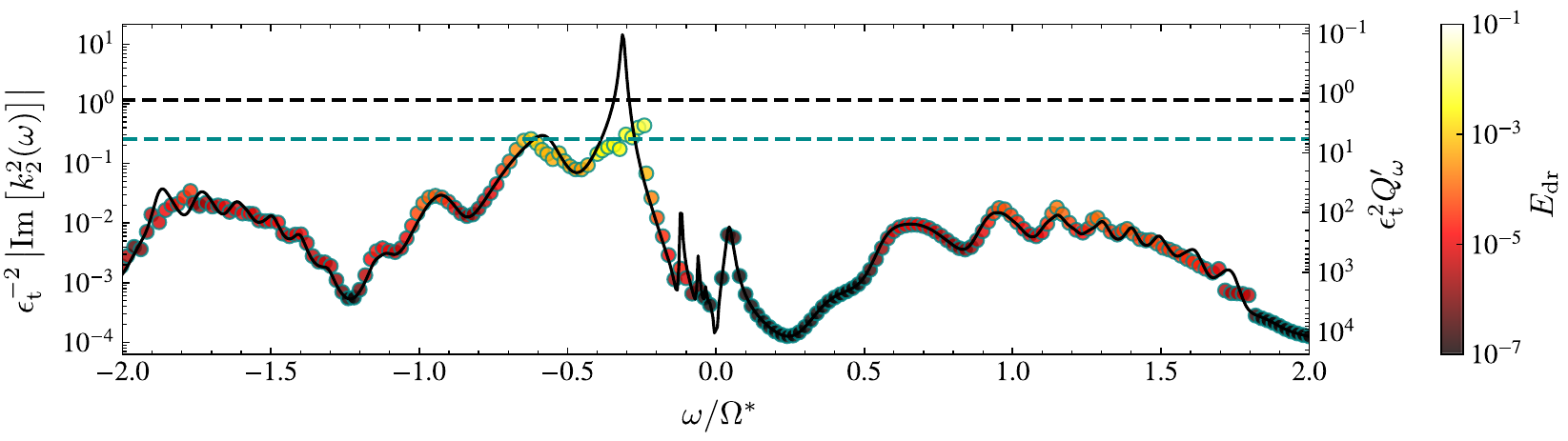}
    \caption{Same as Fig.~\ref{Fig1} but for $\alpha=0.9$ and $\ct=10^{-2}$. $\edr$ is maximal when $\omega=-0.26$ with $\edr=7.8\cdot10^{-3}$ and minimal when $\omega=0.26$ with $\edr=4.3\cdot10^{-10}$.}
    \label{fig:diss0p9}
\end{figure*}
%%%%%%%%%%%%%%%%%%%%%%%%%%%%%%%%%%%%%%%%%%%%%%%%
%%%%%%%%%%%%%
 
It is difficult to predict at a given frequency whether the frequency-dependent dissipation will be attenuated or amplified over linear predictions. The emergence of zonal flows can mitigate the strong activation of intense shear layers, which are associated with the most dissipative peaks, as we observe when comparing the three meridional slices of the radial velocity in Fig.~\ref{fig:ur_omm0p25} at $\omega=-0.25$ (at the peak of linear dissipation for $\alpha=0.7$). The radial velocity amplitude in the waves \textit{decreases} here as $\ct$ is increased. It is not clear whether hidden global modes \citep[also associated with enhanced dissipation in linear theory,][]{LO2021}, still exist in the presence of strong differential rotation. On the other hand, we identify the presence of corotation resonances, for which the Doppler-shifted frequency $\omega-m\delta\Omega$ vanishes \citep[e.g.][]{AP2021}, which appear when zonal flows are strong for $\ct=5\cdot10^{-2}$, particularly in the right panel of Fig.~\ref{fig:ur_omm0p25}. 
We emphasise that the frequency and azimuthal wavenumber in this relationship could differ from  one of the (initial) tidal forcing frequency with $m=2$, since nonlinearities (parametric instabilities or triadic resonances) can excite other modes, such as $m=1$ with lower frequencies, as seen in AB22 \citep[see also][]{BT2018}, which could have their own corotation resonances \citep[reminiscent of][for gravity waves]{Weinberg2012}.

The wave response is very sensitive to $\alpha$ and the initial frequency $\omega$, which dictates both the critical latitude and the angle of IW propagation (hence that of reflection from boundaries). When wave attractors form featuring cyclic behaviour with uniform rotation, energy is strongly focused along these, and IWs can be considered to form modes that vary strongly with $\omega$. With corotation resonances, the precise wave frequency, shell thickness and boundary conditions may matter less, since waves probably travel from critical latitudes where they are launched until they reach a corotation resonance where they are mostly damped \citep[e.g.][]{AP2021}, without being able to form `global modes'. In this `travelling wave' regime \citep[like that for gravity waves approaching critical layers e.g.][]{Z1975,GD1998,BO2010}, we might hypothesise that the dissipation may only depend on the efficiency with which waves are excited, rather than on the precise dissipative mechanism or shape of the container. If so, this might explain the `flattening' of the dissipation observed in Fig.~\ref{fig:dnualpa0p7_neg} when nonlinear effects and $\edr$ are strong. This hypothesis is supported by observing corotation resonances for other initial frequencies than $\omega=-0.25$. 

%%%%%%
Nonlinear effects also mitigate the frequency-averaged tidal dissipation compared with linear predictions, as we observe in Figs.~\ref{Fig1} and \ref{fig:diss0p7}. The strongest reduction is approximately a factor of 3 for $\ct=5\cdot10^{-2}$ and $\Ek=10^{-5}$ (Fig.~\ref{0p7highCt}). 
%63%
For $\alpha=0.9$ (see Fig.~\ref{fig:diss0p9}), appropriate for F-stars or giant planets outside their dynamo regions\footnote{The upper differentially-rotating convective zones of Jupiter and Saturn where zonal flows extend above fractional radii of $\sim0.96$ and $\sim0.86$, respectively \citep[see e.g.][]{GF2022}.}, and $\ct=10^{-2}$, the nonlinear frequency-averaged dissipation is reduced even further from linear predictions (by a factor $4$). This mainly results from the dominant peak of linear dissipation at $\omega=-1/3$ \citep[corresponding to Rossby mode excitation,][]{O2009} being significantly reduced by differential rotation, while for other frequencies with weaker zonal flows nonlinear results mostly follow their linear predictions. We would expect an even larger reduction in the frequency-averaged dissipation for $\ct=5\cdot10^{-2}$ (which are even more demanding cases to resolve spatially and temporally).
%%%%%%%%%%%%%%%%%%%%%%%%%%%%%%%%%%%%%%%%%%%%%
\section{Application to star-planet and binary star systems}
\label{astro}
\subsection{Star-planet systems}
Tidal interactions drive two-body systems towards a stable equilibrium with circular and (equatorially) coplanar orbits, and both rotation periods synchronised with the orbital period \citep[with sufficient angular momentum, otherwise orbital decay towards coalescence is predicted,][]{H1980}. Many close-in planets observed are likely to be synchronised and (mostly) circularised because of strong tidal dissipation in these planets, probably due to IWs \citep[e.g.][]{OL2004,Wu2005,GL2009,Barker2016}. Synchronisation of stellar rotation usually operates on much longer timescales. Damping of IWs in convective envelopes of planet-hosting stars is a possible avenue for tidal dissipation if stars rotate sufficiently rapidly to permit their excitation (requiring $|\omega|\leq 2\Omega$). Two such examples may be WASP-128 and KELT-1 \citep{HT2018,MF2022}, both composed of massive brown dwarfs ($37\,M_\mathrm{J}$ and $27\,M_\mathrm{J}$, respectively) orbiting in $P_\mathrm{o}=2.21$ and $P_\mathrm{o}=1.22$ days around main-sequence stars (of masses $1.16\,M_\odot$ and $1.34\,M_\odot$) with rotation periods $P_\star\approx2.93$ and $P_\star\approx1.52$ days \citep[using][for KELT-1]{EM2021}, respectively. IWs can be tidally-excited in the stellar envelopes since their (asynchronous) tidal frequencies $\omega/\Omega_\star=2(P_\star/P_\mathrm{o}-1)$ are respectively $\omega/\Omega_\star\approx0.65$ and $\omega/\Omega_\star\approx0.5$ (so $|\omega|\leq 2\Omega_\star$). Once these stars become synchronised with their orbits, the companion may subsequently decay on the magnetic braking timescale \citep[e.g.][]{BO2009,DL2015}.

 The stellar tidal amplitudes in WASP-128 and KELT-1 are $\epsilon_\star\approx10^{-4}$ and $\epsilon_\star\approx4\cdot10^{-4}$, respectively, two of the highest\footnote{We expect $\ct\approx\epsilon_\star$ since $k_2\ll1$ given the thin envelopes of main-sequence G and F stars (though $k_2$ could be larger and $O(1)$ during the pre-main sequence).}
in stars hosting planetary companions. For $\ct\ll10^{-2}$ with $\Ek\gtrsim10^{-5}$ \citep[as predicted for turbulent viscosities by mixing-length theory or simulations, e.g.][]{BC2022}, we have shown that nonlinear effects may not significantly modify tidal dissipation in our model (though this statement may not hold for lower viscosities, such as if microscopic rather than turbulent viscosities are relevant for $\Ek$). 
We can nevertheless derive both the linear frequency-dependent and averaged modified tidal quality factors for these systems. We predict $\qpo\approx7\cdot10^4$ for WASP-128 and $\qpo\approx8\cdot10^4$ for KELT-1 using the frequencies above with $\alpha=0.9$ (black solid curve in Fig.~\ref{fig:diss0p9}) and Eq.~(\ref{qfactor}) with $\epOm\approx0.046$  and $\epOm\approx0.12$, respectively, in our homogeneous stellar model. The lower bound for KELT-1 b is constrained from observations to be $Q'\approx2.33^{+0.36}_{-0.38}\cdot10^6$ given that orbital evolution has not been unambiguously detected (as for WASP-128 b). 
%The observed value for $\dot P$ in Hodzic is not very clear! (it implies lower bound is $\qp\sim10^3-10^4$) but no `significant' orbital decay was detected. 
The prediction for frequency-averaged IW tidal dissipation is $Q'\approx 2.6\cdot10^6$ for KELT-1 \citep{B2020,MF2022}, based on a realistic fluid model\footnote{Our model predicts frequency-averaged values (Eq.~(\ref{eq:Lambda1})) $Q'\approx6\cdot10^{2}$ for KELT-1 and $Q'\approx90$ for WASP-128. However, assuming a homogeneous body with a dense core may be less applicable for thin shells ($\alpha=0.9$) due to the strong resonant peak at $\omega\approx-1/3$. Discrepancies with more realistic models are highest for such large $\alpha$ \citep[e.g.~Fig.~7 of][]{O2013}.}, and \cite{HT2018} inferred $Q'\approx10^{7}$ for WASP-128 assuming a dynamical equilibrium where tidal and wind braking torques balance. Thus, operation of IW dissipation is not inconsistent with these observations. 

The tidal amplitude inside planets $\epsilon_\mathrm{p}$ is typically much larger than $\epsilon_\star$ (Fig.~23 of AB22). However, planetary rotation rates $\Omega_\mathrm{p}$ are unknown, usually assumed to be synchronised, i.e.~$\Omega_\mathrm{p}\approx\Omega_\mathrm{o}$ (probably driven by IW dissipation). We have more information on planetary eccentricities $e$, and we can make predictions for when eccentricity tides (i.e.~when $e\ne 0$ drives tidal interactions, for which tidal amplitudes are $\propto e$) may become nonlinear. The frequency for a given ($n,m$) tidal component is $\omega=n\Omega_\mathrm{o}-m\Omega_{p}$ where $n$ labels temporal harmonics of the orbital motion, and $(m,n)=(2,3)$ is usually dominant \citep[but $(m,n)\in\{(2,1),(0,1)\}$ can also contribute,][]{O2014}. Assuming $\Omega_\mathrm{p}=\Omega_\mathrm{o}$, we have $\omega/\Omega_\mathrm{p}=1$ ($-1$ with $n=1$). We have searched the \url{exoplanet.eu} database for planets satisfying $7e\,\ct/2>10^{-2}$, assuming $k_2=0.3$ as a lower bound\footnote{Based on constraints for the Hot Jupiters HAT-P-13 b \citep[$k_2=0.31^{0.08}_{-0.05}$,][]{BK2016}, WASP-121 b \citep[$k_2=0.39\pm0.8$,][]{HC2019}, and WASP-103 b \citep[$k_2=0.59^{+0.45}_{-0.53}$,][]{BA2022}.} \citep[see also][for $k_2$ computations in fast rotators]{DL2022},
 %,and for Jupiter associated to Io's tide $k_2=0.565\pm0.006$ \citep{GF2022}. 
which selects HAT-P-65 b, HAT-P-67 b, HATS-18 b, and HATS-24 b. These are inflated planets cf. Jupiter and $e\gtrsim 0.1$, 
%thin radiative envelope \citep[e.g.][]{TG2019}.
which rotate at approximately $10\%$ of their breakup velocities (assuming synchronism, i.e.~$\epOm\approx0.1$) and tidal amplitudes $7e\,\ct/2$ as large as $10^{-2}$ for HATS-24 b, HATS-18 b, HAT-P-67 b, and closer to $5\cdot10^{-2}$ for HAT-P-65 b. Nonlinear IW effects are likely to be even more important (with larger $\ct$) for the past tidal evolution of Hot Jupiter systems, since young planets have larger radii, probably rotate more rapidly, and may not have been tidally synchronised or circularised. Table \ref{tab:qp_hj} provides both the linear and nonlinear predictions for $Q'$ (frequency-dependent and averaged) for these systems, for various $\alpha$ as planetary internal structures are uncertain. Varying $\alpha$ leads to important numerical differences in $Q'$, which span the range from $10^2$ to $10^5$, with even higher values for a tiny core ($Q'>10^7$ for $\alpha=0.1$). However, the linear and nonlinear frequency-dependent estimates are often comparable, with moderate discrepancies when $\ct=5\cdot10^{-2}$. The frequency-averaged $Q'$ is generally smaller than the frequency-dependent value, except close to resonant peaks. This is because the former is dominated by resonant peaks in linear theory. The strong dependence on $\alpha$ suggests the possibility of constraining planetary structures from observations inferring their tidal evolution.

\begin{table}\centering
\begin{tabular}{c|c|c|c|c}
$\alpha$ & $\qpo$ (l) & $\qpo$ (nl) & $\qp$ (l) & $\qp$ (nl) \\
\hline
0.3 & $10^5$ & $[10^5,2\cdot10^5]$ & $5\cdot10^4$ & $[5\cdot10^4,5\cdot10^4]$ \\
0.5 & $7\cdot10^4$ & $[7\cdot10^4,8\cdot10^4]$ & $3\cdot10^3$ & $[4\cdot10^3,5\cdot10^3]$ \\
0.7 & $3\cdot10^3$ & $[3\cdot10^3,10^4]$ & $6\cdot10^2$ & $[8\cdot10^2,10^3]$\\
0.9 & $8\cdot10^{3}$ & $7\cdot10^{3}$ & $8\cdot10^1$ & $3\cdot10^2$\\
\end{tabular}
    \caption{Tidal quality factors $\qpo$ (frequency-dependent) and $\qp$ (frequency-averaged) for eccentricity tides in Hot Jupiters. % (listed below). 
    We compute linear (l) and nonlinear (nl) estimates assuming $k_2=0.3$ and $\epOm=0.1$. Values in brackets are for $\ct=[10^{-2},5\cdot10^{-2}]$ (otherwise $\ct=10^{-2}$).}
\label{tab:qp_hj}
\end{table}

\subsection{Application to late-type eclipsing binaries}
\label{apA}
There is strong evidence for tidal circularisation and synchronisation in late-type binaries \citep[e.g.][]{MM2005,MM2006}. Equilibrium tides are not believed to be sufficient to explain circularisation of old binaries\footnote{See \cite{T21} for arguments to the contrary.} \citep[][]{ZB1989,MM2004,Z2013}, but dissipation of IWs could explain observations \citep{B2022}. Regarding tidal synchronisation, \cite{LV2017} reported rotation periods for several hundred low-mass stars in eclipsing binaries (EBs). 
The majority are approximately synchronised, but a substantial fraction 
rotate sub-synchronously, possibly because they feature differential rotation or because synchronisation is ongoing. We calculate $Q'$ for the asynchronous tide in these systems due to IW dissipation in their convective envelopes.  
We choose EBs satisfying $P_\mathrm{o}>P_1/2$ (for IW excitation), where $P_1$ is the primary's rotation period \citep[i.e.~the more massive star, using $P_\mathrm{1\,min}$ in Table 2 of][]{LV2017,PP2022}. To compute $\epsilon$ and $\epOm$, we also use data in \cite{WA2019} (i.e.~masses, radii, and ages). We enforce $\epsilon>10^{-2}$, inferring $a$ using Kepler's $3^\text{rd}$ law, and list the selected Kepler EBs in Table~\ref{tab:Ebs}. $M_1$ is rounded to the nearest tenth to match stellar models \citep[computed with STAREVOL,][]{AP2019,AM2019} to infer both $\alpha$ (rounded to the nearest value in \S~\ref{scan}) and $k_2=3\rho/(5\overline{\rho}-3\rho)$ \citep[see][for a similar homogeneous fluid model]{O2013} at a given age, with $\overline{\rho}$ the mean stellar density and $\rho$ the density in the middle of the envelope. The (non-)linear $\qpo$ is computed using Eq.~(\ref{qfactor}) with $\omega/\Omega=2(P_1/P_\mathrm{o}-1)$, taken from our closest model ($\alpha,C_t$). The (non-)linear frequency-averaged $Q'=3/(2\Lambda)$ is computed using Eq.~(\ref{eq:Lambda1}) (Eq.~(\ref{eq:Lamb})). We notice  significant discrepancies, sometimes by several orders of magnitude (e.g.~the first system), between frequency-dependent and frequency-averaged values. The moderate $\ct$ in these systems prevents nonlinear effects from significantly impacting quality factors (for the $\mathrm{E}$ studied). The dissipation is very heterogeneous depending on the system, with $1.7<\log_{10}Q'_{\omega}<6$, typically much lower than the frequency-averaged predictions in \citet[Fig.~1]{B2022}, where $Q'\approx 10^7(P_1/10 \,\mathrm{d})^2$ for main-sequence $0.2-1.2M_\odot$ stars. This difference is partly related to the density variation throughout the body allowed in \citet{B2022}, and the adoption of a rigid core here, both of which are important assumptions to relax in future simulations.
\begin{table*}
    \centering
    \begin{tabular}{c|c|c|c|c|c|c|c|c|c|c|c|c}
        System & $\epsilon$ & \multirow{2}*{$M_1$} & \multirow{2}*{age} & \multirow{2}*{$\alpha$} & \multirow{2}*{$\omega$} & \multirow{2}*{$k_2$} & \multirow{2}*{$\epOm$} & \multirow{2}*{$\qpo$ (l)} & \multirow{2}*{$\qpo$ (nl)} & \multirow{2}*{$\qp$ (l)} & \multirow{2}*{$\qp$ (nl)} & \multirow{2}*{$Q_\mathrm{A}'$}\\
        (KIC) & $(\times10^{-2})$ & & & & & & & & & & & \\
        \hline
        10257903 & $1.7$ & $1$ & $7.1\cdot10^9$ & $0.7$ & $0.27$ & $2\cdot10^{-2}$ & $0.18$ & $10^6$ & $10^6$ & $3\cdot10^2$ & $4\cdot10^2$ & $10^5$ \\
        9020426 & $1.3$ & $1$ & $7.8\cdot10^6$ & $0.5$ & $-0.3$ & $0.4$ & $0.19$ & $6\cdot10^3$ & $7\cdot10^3$ & $8\cdot10^2$ & $10^3$ & $6\cdot10^3$\\
        8155368 & $2.4$ & $0.8$ & $10^{10}$ & $0.7$ & $-0.94$ & $0.2$ & $0.46$ & $5\cdot10^1$ & $[5\cdot10^1,2\cdot10^2]$ & $3\cdot10^1$ & $[5\cdot10^1,8\cdot10^1]$ & $10^4$ \\
        7985167 & $1.5$ & $1.3$ & $3.5\cdot10^9$ & $0.9$ & $-0.72$ & $10^{-3}$ & $0.28$ & $3\cdot10^2$ & $2\cdot10^2$ & $2\cdot10^1$ & $7\cdot10^1$ & $8\cdot10^4$ \\
        7885570 & $1.7$ & $1.5$ & $2.3\cdot10^9$ & $0.9$ & $-0.52$ & $2\cdot10^{-4}$ & $0.29$ & $10^2$ & $2\cdot10^2$ & $2\cdot10^1$ & $7\cdot10^1$ & $2\cdot10^5$ \\
%        KIC 6464285 & $1.1$ & $3.8\cdot10^9$ & $0.9$ \aava{closer to $0.8$} & $-0.19$ &  &  &  &  &  &  \\
        6311637 & $1.4$ & $1.4$ & $3.5\cdot10^9$ & $0.9$ & $-0.66$ & $3\cdot10^{-4}$ & $0.30$ & $10^2$ & $7\cdot10^1$ & $10^1$ & $7\cdot10^1$ & $10^5$\\
        6283224 & $1.2$ & $1$ & $5.3\cdot10^9$ & $0.7$ & $-1.24$ & $10^{-2}$ & $0.42$ & $6\cdot10^4$ & $6\cdot10^4$ & $5\cdot10^1$ & $8\cdot10^1$ & $9\cdot10^3$ \\
        3662635 & $1.2$ & $1.3$ & $5.6\cdot10^8$ & $0.9$ & $0.04$ & $2\cdot10^{-4}$ & $0.15$ & $10^4$ & $10^4$ & $6\cdot10^1$ & $3\cdot10^2$ & $10^6$\\
        3344427 & $1.3$ & $0.9$ & $6.7\cdot10^9$ & $0.7$ & $-0.77$ & $3\cdot10^{-2}$ & $0.27$ & $6\cdot10^3$ & $6\cdot10^3$ & $10^2$ & $2\cdot10^2$ & $2\cdot10^4$ \\
        2447893 & $1.5$ & $0.9$ & $1.4\cdot10^7$ & $0.5$ & $0.21$ & $0.3$ & $0.16$ & $3\cdot10^5$ & $8\cdot10^5$ & $10^3$ & $2\cdot10^3$ & $5\cdot10^4$\\
    \end{tabular}
    \caption{Frequency-dependent and averaged modified tidal quality factors $\qpo$ and $\qp$ from linear (l) and nonlinear (nl) simulations for several EBs. Nonlinear values in brackets are for $\ct=[10^{-2},5\cdot10^{-2}]$ (otherwise $\ct=10^{-2}$). The last column shows approximate frequency-averaged linear predictions from \citet{B2022} accounting for realistic stellar density profiles: $Q_\mathrm{A}'\approx10^7(P_1/10 \,\mathrm{d})^2$ for main-sequence $0.2-1.2M_\odot$ stars or their Fig.~1 (top panel). This can differ substantially from our predictions for the frequency-averaged value in particular (less so for $\qpo$).
    }
    \label{tab:Ebs}
\end{table*}
%%%%%%%%%%%%%%%%%%%%%%%%%%%%%%%%%%%%%%%%%%%%
\section{Conclusions}\label{conc}
We have simulated tidally-forced inertial waves in hydrodynamical spherical shell models of convective envelopes in rotating low-mass stars and giant planets. Our main goal was to determine the validity of linear theory, and particularly the widely-applied frequency-averaged inertial wave dissipation, in modelling tidal dissipation in stars and planets.

Our nonlinear simulations have demonstrated that the strongly frequency-dependent dissipation predicted by linear theory is increasingly smoothed out by nonlinearities for increasing tidal amplitudes (larger $\ct$), for thinner shells (larger $\alpha$), and for smaller viscosities (smaller $\mathrm{E}$). Our results predict tidal energy transfer rates that differ from linear predictions by up to 3 orders of magnitude at a given frequency due to the generation of differential rotation and its back-reaction on the waves. However, we have found the frequency-averaged prediction from linear theory to be more robust, typically predicting the frequency-averaged nonlinear dissipation to within a factor of $4$. The largest disagreements are found where nonlinearities are strongest, and the strongest peaks of linear dissipation are substantially attenuated. 

This suggests that the frequency-averaged linear theory predictions for inertial wave dissipation are relatively robust even when including nonlinear effects, and thus may be reasonable to apply to modelling tidal evolution of stellar and planetary statistical populations. However, we caution that tidal evolution in a given system (at a given time and tidal frequency) might differ substantially. 
It may also greatly change when taking into account realistic density profiles, especially for thin shells, and the influence of a radiative interior  -- here modelled as a rigid core, so assumed to be strongly stratified as expected in low-mass stars. Otherwise, when the buoyancy frequency at the radiative/convective interface is comparable to rotation \citep[as expected for Jupiter and Saturn, e.g.][]{MF2021}, tidal dissipation from inertial waves could be mitigated by the presence of a stably-stratified dilute core \citep[for Jupiter,][Dhouib et al. submitted]{D2023,L2023} or on the contrary enhanced (for Saturn, Pontin et al. submitted). Future work should explore the sensitivity of boundary conditions and more realistic planetary/stellar structures on (non-)linear simulations and how well using the frequency-averaged value models tidal evolution in real frequency-dependent systems.

The ``smoothing-out" of the dissipation predicted by linear theory is reminiscent of the transition from global modes (with strongly frequency-dependent dissipation) to travelling waves (with much weaker dependence on frequency) for tidally-excited gravity waves in stellar radiation zones. We have identified the key role of corotation resonances (where the Doppler-shifted wave frequency vanishes) that may explain this tendency in our simulations of tidally-generated differential rotation. It is essential to explore further the role of turbulent convection and convectively-generated differential rotation. Tidal generation of differential rotation may compete with convection in the closest binaries or in Hot Jupiters. Our work thus motivates further studies of tidal inertial waves and of their interactions with differential rotation, turbulent convection and magnetic fields using more realistic density profiles. Finally, it would be worth exploring models permitting realistic nonlinear couplings between wavelike and non-wavelike tidal flows. 

%Authors are required to provide line numbering in the manuscript. Line numbering makes it easier for the review to references specific places in the manuscript.
%This functionality has been built into AASTeX since v6.0.  The {\tt\string linenumbers} style option invokes the lineno style file to number each article line in the left margin.
% anonymous option

%% Also note that the akcnowlodgment environment does not support long amounts of text. If you have a lot of people and institutions to acknowledge, do not use this command. Instead, create a new \section{Acknowledgments}.
\section*{acknowledgments}
  Funded by STFC grants ST/S000275/1 and ST/W000873/1, and by a Leverhulme Trust Early Career Fellowship to AA. Simulations were undertaken on the DiRAC Data Intensive service at Leicester, operated by the University of Leicester IT Services, which forms part of the STFC DiRAC HPC Facility (\href{www.dirac.ac.uk}{www.dirac.ac.uk}). The equipment was funded by BEIS capital funding via STFC capital grants ST/K000373/1 and ST/R002363/1 and STFC DiRAC Operations grant ST/R001014/1. DiRAC is part of the National e-Infrastructure. This research has made use of data obtained from or tools provided by the portal exoplanet.eu of The Extrasolar Planets Encyclopaedia, and of NASA’s Astrophysics Data System Bibliographic Services. We thank M. Rieutord, R. V. Valdetarro and C. Baruteau for providing the LSB code and A. Guseva and the referee for helpful comments.
%\end{acknowledgments}


%\vspace{5mm}
\facilities{CDS, ADS}


\software{MagIC dynamo code (version 5.10) at https://magic-sph.github.io/\\
STAREVOL \citep{AP2019}\\
LSB \citep{RV2010}
}



\appendix
\section{Discussion on tidally-induced zonal flows}
\subsection{Tidal forcing amplitude dependence}
\label{scal}
\begin{figure}
    \centering
    \includegraphics[width=0.49\textwidth]{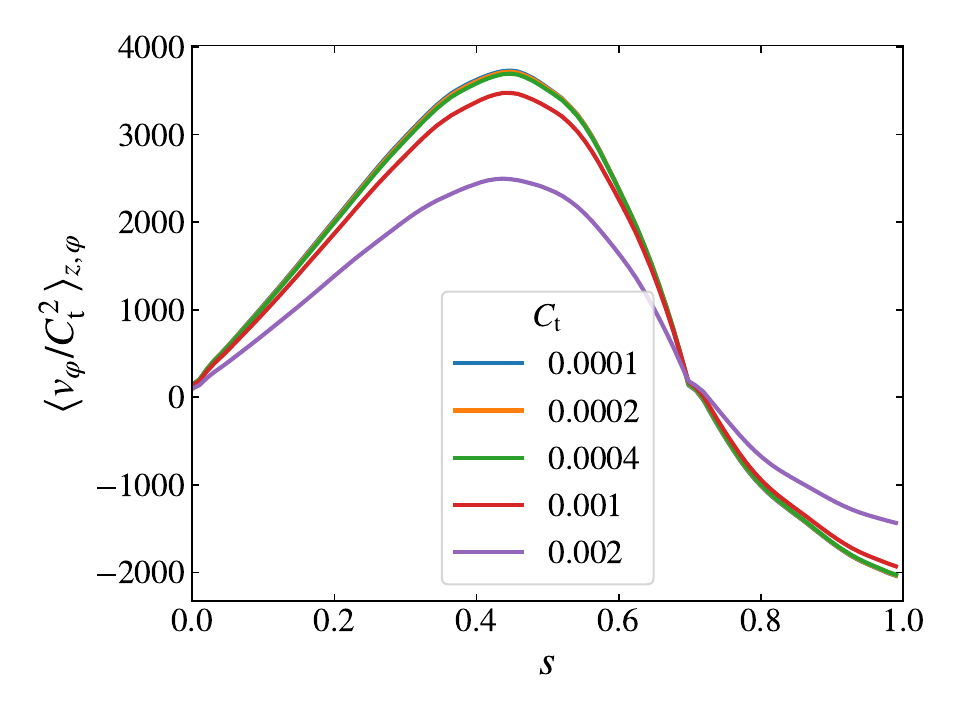}
    \includegraphics[width=0.49\textwidth]{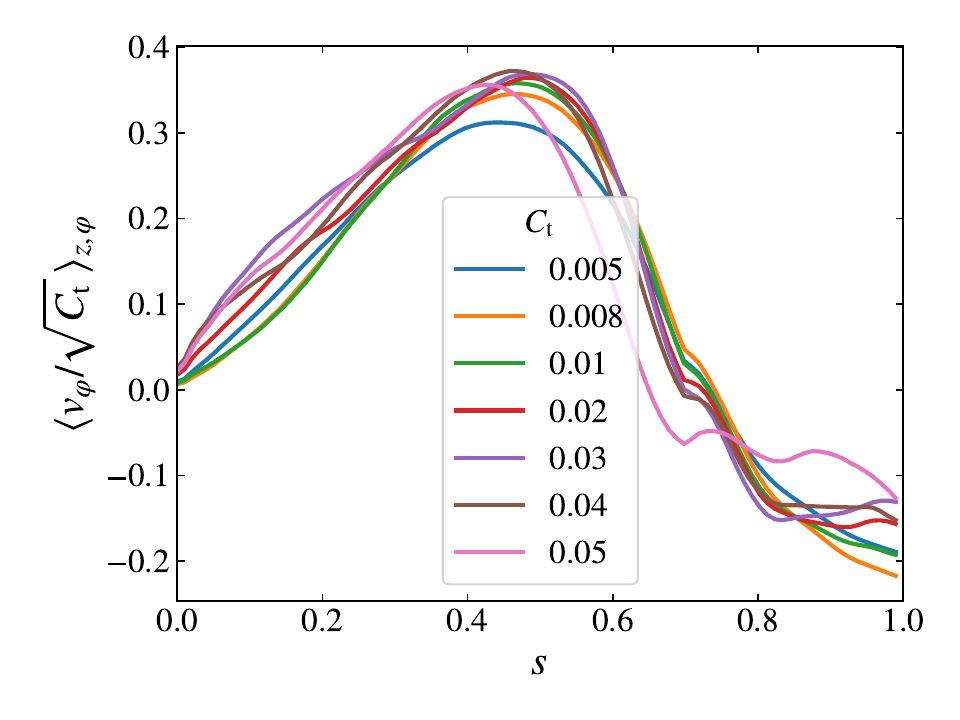}
    \caption{Azimuthal and vertical average of the azimuthal velocity normalised by the tidal amplitude squared $\langle v_\varphi/\ct^2\rangle_{z,\varphi}$ (\textit{left}, or by $\langle v_\varphi/\sqrt{\ct}\rangle_{z,\varphi}$ (\textit{right}), against the distance to the rotation axis $s$, for low (\textit{left}) and moderate to high (\textit{right}) values of the tidal amplitude $\ct$ in different colours. The tidal frequency, core size, and Ekman number are set to $\omega=-0.25$, $\alpha=0.7$, and $\Ek=10^{-5}$, respectively in both panels.
    }
    \label{profil}
\end{figure}
\begin{figure}
    \centering
    \includegraphics[width=0.49\textwidth]{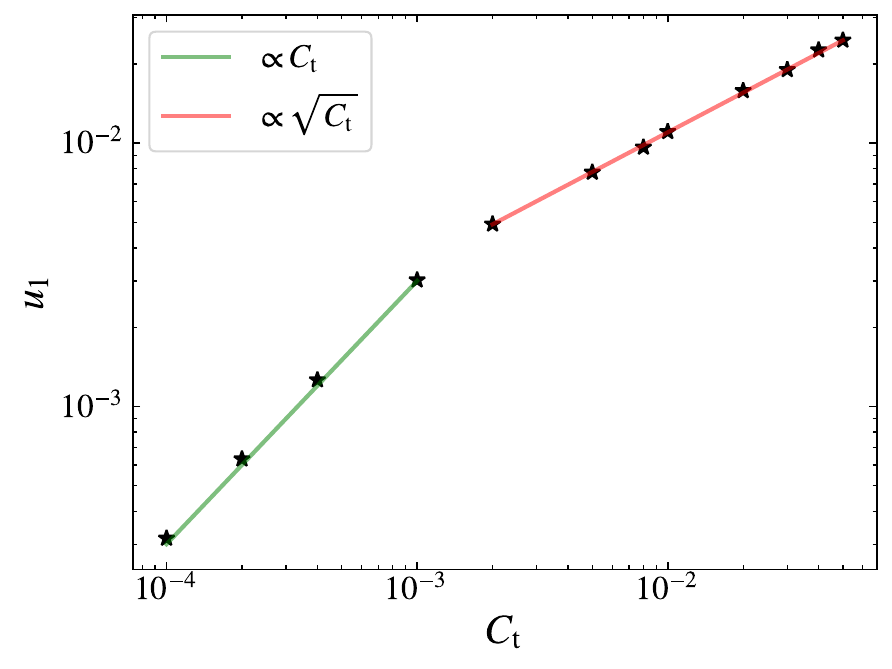}
    \includegraphics[width=0.49\textwidth]{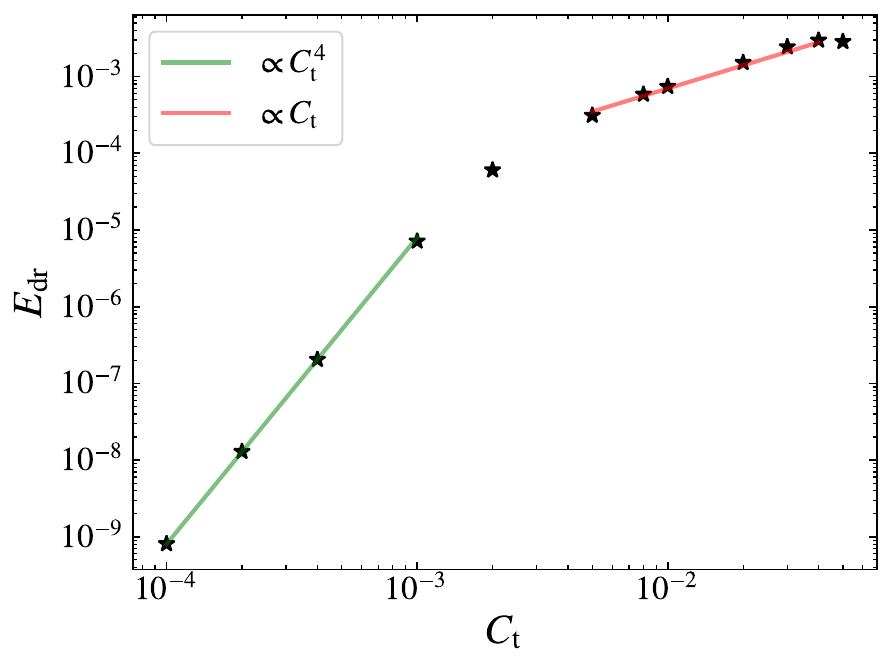}
    \caption{\textit{Left:} Non-axisymmetric component of the tidal flow magnitude $u_1$ against the tidal amplitude $\ct$, when filtering out the zonal flow $m=0$ component from the kinetic energy spectrum summed over $m\neq0$ components.  
    \textit{Right:} Energy in the differential rotation $\edr$ against tidal amplitude $\ct$. The different nonlinear regimes are indicated with coloured lines in both panels. The tidal frequency, core size, and Ekman number are set to $\omega=-0.25$, $\alpha=0.7$, and $\Ek=10^{-5}$, respectively.}
    \label{edr}
\end{figure}

The structures of the zonal flows generated by IW
nonlinear interactions are difficult to predict as they are strongly dependent on the initial tidal frequency, core size and Ekman number, which dictate the shear layer structure and the locations of strong dissipation inside the shell. Though the energy inside the differential rotation typically increases with decreasing Ekman number, the nature of this dependence varies with frequency in a non-trivial way \citep[see Fig.22 of][]{AB2022}. In Figs.~\ref{profil} and \ref{edr}, we explore the zonal flow strength's dependence on tidal amplitude for a fixed tidal frequency, core size and Ekman number. Curiously, two regimes clearly emerge depending on the strength of nonlinear effects. In the weakly nonlinear regime for $\ct\lesssim10^{-3}$, we can anticipate this regime following \cite{T2007,AB2022}. Using the notations  $u_1$ for the linear tidal flow and $u_2$ for the induced zonal flow, the scaling law  $u_2\propto u_1^2\propto C_t^2$ is directly obtained from balancing zonal flow viscous dissipation $\Ek\Delta\bm u_2$ with the nonlinear advection term $(\bm u_1\cdot\bm\nabla)\bm u_1$ which forces the zonal flow. This regime is evidenced for $\ct\lesssim10^{-3}$ in Fig.~\ref{profil} for the zonal flow profile, and in Fig.~\ref{edr} (green scaling), where we show that $u_1\propto C_\mathrm{t}$ (time derivative and/or the Coriolis force acting on $u_1$ balances tidal forcing) and $\edr\propto u_2^2\propto C_\mathrm{t}^4$. Note that both the zonal flow and tidal inertial wave length-scales are independent of $C_\mathrm{t}$ (as shown in Fig.~\ref{profil} for the zonal flow). In the moderately nonlinear regime for $2\cdot10^{-3}\lesssim  C_\mathrm{t}<5\cdot10^{-2}$, $u_1$ is proportional to $\sqrt{\ct}$ (as is $u_2$), as we show in Fig.~\ref{profil} (right panel) and Fig.~\ref{edr} (left panel, red scaling), making $\edr\propto u_2^2\propto C_\mathrm{t}$ (right panel, red scaling). Such a scaling can be obtained if the waves satisfy a balance between nonlinear advection $(\bm u_1\cdot\bm\nabla)\bm u_1$ and tidal forcing, and then if the axisymmetric component of $(\bm u_1\cdot\bm\nabla)\bm u_1$ balances $(\bm u_2\cdot\bm\nabla )\bm u_2$. For the strongly nonlinear regime when $\ct\geq5\cdot10^{-2}$, $\edr$ seems to saturate (right panel of Fig.~\ref{edr}), possibly because hydrodynamical instabilities limit zonal flow strengths to become independent of the tidal forcing amplitude.

\subsection{Time to generate zonal flows}
\label{time}
%%%%%%%%%%%%
%%%%%%%%%%%%
\begin{figure}
    \centering
    \includegraphics[width=0.5\textwidth]{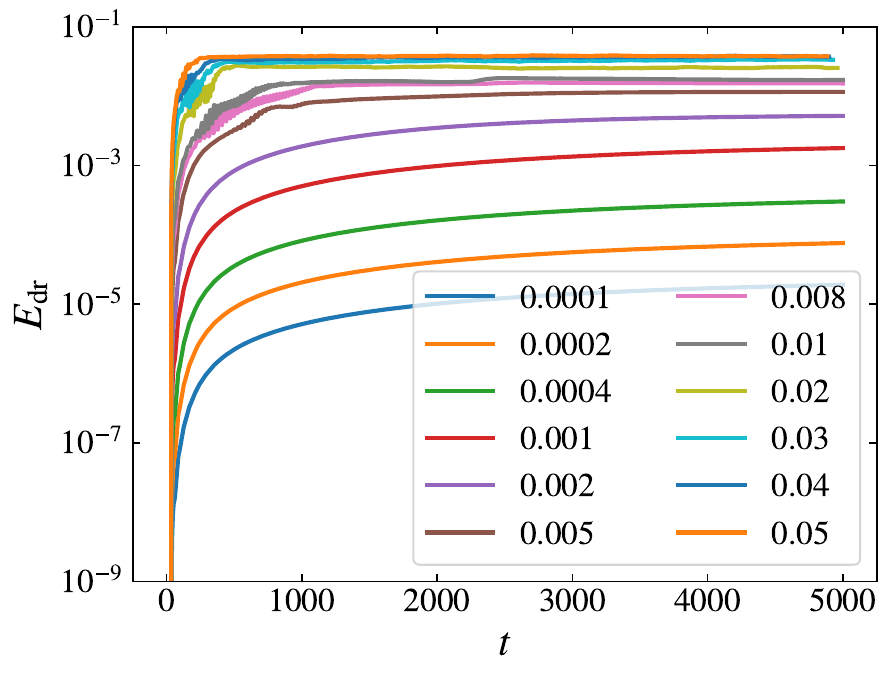}
    \caption{Energy in the differential rotation $\edr$ against time for different tidal forcing amplitudes $\ct$ shown in different colours.  The tidal frequency, core size, and Ekman number are set to $\omega=-0.25$, $\alpha=0.7$, and $\Ek=10^{-5}$, respectively.}
    \label{setin}
\end{figure}
For all simulations, we assume their typical timescales are fast compared with tidal evolutionary timescales so that we can treat the equilibrium tide as a prescribed (not secularly evolving) time-dependent flow. Given that resonant peaks of tidal dissipation may induce fast changes in the dynamical evolution of orbits and rotations and hence the tidal frequency, %\citep[as evidenced in][for r-modes; Kwok et al. in prep. for inertial waves]{PS2023}  %AJB: personally I don't like too many in prep papers and this idea isn't original to PS2023? Don't feel strongly though...
unless coincident changes in internal structure or rotation evolve to maintain resonances \citep[e.g.][]{WS2002,FL2016,LC2020}, one can wonder if the development of zonal flows at a peculiar resonant tidal frequency (i.e.~for a specific configuration of the system) are fast enough compared to tidal evolution timescales for our results regarding them to be valid. %In the first aforementioned papers, the time spent at resonances is numerically found to be in the range of $10^3-10^4$ years. 
In our simulations, we observe that the energy in the differential rotation reaches a steady state after a few thousand $\Omega^{-1}$ units \citep[as shown in Fig. \ref{setin}, also in][]{AB2022}. Indeed, for low amplitudes, a saturated state is reached when the rate of local angular momentum deposition into the mean flow by the waves balances the (presumably turbulent in reality) viscous damping rate for the mean flow. This rate is also dependent on $\ct$, and proportional to $u_1^2$, so the zonal flow is faster to develop for higher tidal amplitude forcing (see Appendix \ref{scal} and Fig. \ref{setin}).
When estimating the zonal flow lengthscale to be a few tenths of the radius of the body, for instance $l/R\sim0.2$ (e.g.~Fig.~\ref{profil}), the corresponding viscous timescale is $\tau_\nu=l^2/\mathrm{Ek}\sim4000\Omega^{-1}$ for $\Ek=10^{-5}$ (even shorter for high $\ct$ as evidenced in Fig. \ref{setin}). For a fast one-day or ten-day rotating young star, this means timescales of approximately 10--100 years. This time is short, and probably shorter than tidal evolutionary timescales. As a result, it is plausible that tidally-induced differential rotation has enough time to set in before the system evolves out of the resonance -- and more globally for every tidal frequency not in a resonant peak. We stress that this statement may only be valid when using mixing-length theory (large) estimates for the viscosity. (On the other hand, if we apply tiny values of atomic viscosity involving $\Ek=10^{-12}$ or lower, viscous forces may be too weak for such a balance to be attained before the system has evolved through a resonant peak.) To conclude, we find that tidal generation of zonal flows can be very rapid for large $\ct$ (even a few hundred rotation times for large $\ct$, since the angular momentum deposition is faster, as evidenced in Fig. \ref{setin}), and due to the localised nature of the differential rotation produced this can happen prior to changing the bulk rotation of the body \citep[see the related problem of gravity waves in stars, e.g.][]{Guo2023}. 
%%%%%%%%%%%%%%%%%%%%%%%%%%%%%%%%%%%%%%%%%%%%%%%%%%%%%%%%%

%Authors may use the online length calculator to get an estimate of the number of word and float quanta in their manuscript. The calculator is located at \url{https://authortools.aas.org/Quanta/newlatexwordcount.html}.

%A handy "cheat sheet" that provides the necessary \latex\ to produce 17 different types of tables is available at \url{http://journals.aas.org/authors/aastex/aasguide.html#table_cheat_sheet}.


\bibliography{biblio}{}
\bibliographystyle{aasjournal}



\end{document}